\pdfoutput=1

\documentclass[aps,twocolumn,amsmath,amssymb,preprintnumbers,floatfix,prb,superscriptaddress,longbibliography]{revtex4-2}

\usepackage[utf8]{inputenc}
\usepackage{newtxtext}
\usepackage[upint]{newtxmath}
\usepackage{microtype}
\usepackage{textcomp}
\usepackage{eucal}
\usepackage{bm}
\usepackage{siunitx}
\usepackage{comment}

\usepackage{enumerate}
\usepackage{amsfonts}
\usepackage{amsmath}
\usepackage{amssymb}
\usepackage{color}
\usepackage{soul}

\usepackage{graphicx}

\usepackage[colorlinks,allcolors=blue]{hyperref}
\usepackage[capitalize]{cleveref}

\definecolor{DarkRed}{rgb}{0.65,0,0}%
\definecolor{Green}{rgb}{0,0.3,0.3}
\definecolor{Purple}{rgb}{0.3,0,0.65}
\definecolor{Red}{rgb}{1,0,0}
\definecolor{Blue}{rgb}{0,0,0.85}
\definecolor{Magenta}{rgb}{1,0,1}

\newcommand{\ve}[1]{\boldsymbol{#1}}

\newcommand{\vk}{{\ve{k}}} 

\newcommand{\vecx}{\hat{\boldsymbol{x}}} 
\newcommand{\vecy}{\hat{\boldsymbol{y}}} 
\newcommand{\vecz}{\hat{\boldsymbol{z}}} 

\newcommand{\e}[1]{\mathrm{e}^{#1}}

\newcommand{\vecsigma}{\boldsymbol{\sigma}}

\newcommand{\vecn}{\ve{n}}
\newcommand{\vecd}{\ve{d}}
\newcommand{\veca}{\ve{a}}
\newcommand{\vecb}{\ve{b}}

\newcommand{\vecm}{\ve{m}}

\newcommand{\vecH}{\ve{H}}

\newcommand{\vecM}{\ve{M}}

\newcommand{\eg}{\textit{e.g. }}

\def\ii{\mathrm{i}}

\newcommand{\be}{\begin{equation}}
\newcommand{\ee}{\end{equation}}

\newcommand{\prlsection}[1]{\textit{#1}.\kern0.05em---\kern0.05em\ignorespaces}

\begin{document}
\title{Giant magnetoanisotropy in the Josephson effect and switching of staggered order in antiferromagnets}
\author{Vemund Falch}
\affiliation{Center for Quantum Spintronics, Department of Physics, Norwegian \\ University of Science and Technology, NO-7491 Trondheim, Norway}
\author{Jacob Linder}
\email[Corresponding author: ]{jacob.linder@ntnu.no}
\affiliation{Center for Quantum Spintronics, Department of Physics, Norwegian \\ University of Science and Technology, NO-7491 Trondheim, Norway}

\begin{abstract}
We predict that the amplitude of the Josephson current through an antiferromagnetic weak link changes by several orders of magnitude upon rotation of the Néel order parameter characterizing the staggered magnetic order. This occurs due the presence of spin-orbit coupling arising from structural inversion asymmetry which makes the band gap in the antiferromagnet highly sensitive to the staggered order parameter direction. We also demonstrate that when phase-biasing the junction, magnetization dynamics is induced which switches the direction of the Néel vector in the antiferromagnet. These results reveal an interesting versatility of antiferromagnetic Josephson junctions as they offer both a large tunability of the supercurrent magnitude via the staggerered magnetization and phase-coherent control over the Néel order parameter. 
\end{abstract}
\maketitle
\section{Introduction} 
In recent years, it has been realized that antiferromagnetic materials can give rise to effects in spin-electronics which are potentially useful technologically as well as interesting from a fundamental physics viewpoint \cite{baltz_rmp_18}. This includes ultrafast magnetization dynamics on the GHz scale \cite{macneill_prl_19} as well as controllable motion of antiferromagnetic domain walls \cite{manchon_rmp_19}. The use of superconducting materials in spintronics \cite{linder_nphys_15} has also been shown to give rise to  phenomena such as strongly enhanced spin-accumulation \cite{hubler_prl_12, bergeret_rmp_18} and spin Hall effects compared to what is possible in the normal state of metals \cite{wakamura_natmat_15}. 

Driven by these findings, heterostructures of antiferromagnetic and superconducting materials has very recently also attracted attention \cite{rabinovich_prr_19, jakobsen_prb_20, johnsen_prb_21, bobkov_prb_21, fyhn_pr_22}. In particular the case of uncompensated antiferromagnetic interfaces have been predicted to give rise to interesting spin-dependent effects in superconductors \cite{andersen_prb_05}, although compensated interfaces have also been shown to offer spin-functionality \cite{johnsen_prb_21}. Moreover, when spin-orbit interactions are present due to \eg structural inversion asymmetry breaking, additional phenomena emerge such as tunable quantum phase batteries \cite{rabinovich_prr_19}. 

In a broader perspective, multiple spontaneous symmetry breaking in condensed-matter systems ranks among the most profound emergent phenomena
in many-body physics. It is a topic that is not only of interest in terms of studying properties of specific condensed-matter systems, but also because spontaneous breaking of symmetry is responsible for a wide range of physical effects, including the mass differences of elementary particles,
phase transitions in condensed matter systems, and
emergent phenomena in biology \cite{anderson_textbook_80}. 

\begin{figure}[t!]
    \centering
    \includegraphics[width=\columnwidth]{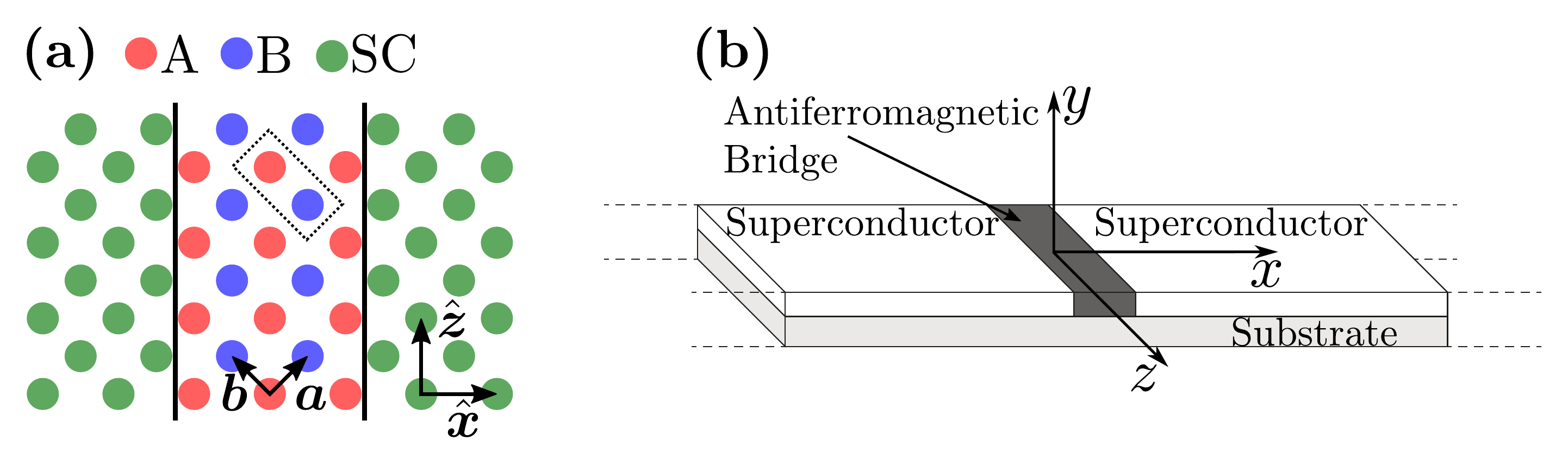}
    \caption{(Color online) (a) The geometry of the (110) junction with nearest neighbor unit vectors $\veca, \vecb$ of length $a$ and basis vectors $\vecx, \vecz$. The unit cell is shown with a dotted square and contains one site from each sublattice. (b) Proposed experimental setup for a thin Josephson junction deposited on a substrate which breaks structural inversion symmetry, allowing for Rashba-type spin-orbit coupling. The direction of inversion symmetry breaking is along the $\vecy$-axis whereas the staggered order parameter $\vecm=(\vecm^A - \vecm^B)/2$ inside the antiferromagnetic region can point in an arbitrary direction. Here, $\vecm^{A,B}$ is the magnetization on sublattice A and B.
		}
    \label{fig:model}
\end{figure}

Here, we investigate the mutual interaction between the supercurrent magnitude and the Néel order parameter in a Josephson junction comprised of two superconductors and an antiferromagnet (Fig. \ref{fig:model}). We discover a giant anisotropy in the Josephson effect: by rotating the Néel order parameter, the supercurrent changes by several orders of magnitude. The underlying physics is explained by the momentum-dependence of the band gap in a spin-orbit coupled antiferromagnet. Moreover, we demonstrate that when phase-biasing the junction via \eg external flux, magnetization dynamics is induced which switches the direction of the Néel vector in the antiferromagnet. This occurs even for extremely small magnetic fluxes which have a negligible Zeeman-coupling to the sublattice magnetization and induce a negligible Meissner response in the superconductors. The results described above show that antiferromagnetic Josephson junctions offer a magnetic transistor-like functionality with strongly tunable supercurrents and that they can be used to both induce magnetization dynamics of the Néel order in addition to detecting its orientation.

\section{Theoretical framework} 
The starting point for computing the supercurrent in an antiferromagnetic Josephson junction with spin-orbit coupling is the following mean-field Hamiltonian:
\begin{align}
H &= -t\sum_{\langle i,j \rangle \sigma} c_{i\sigma}^\dag c_{j\sigma} - \mu\sum_{i\sigma} c_{i\sigma}^\dag c_{i\sigma} + \sum_{i\alpha\beta} \vecm_i \cdot c_{i\alpha}^\dag \vecsigma_{\alpha\beta} c_{i\beta} \notag\\
&-\frac{\ii V_R}{a} \sum_{\langle i,j\rangle\alpha\beta} [\hat{\vecn} \cdot (\vecsigma \times \vecd_{ij})]_{\alpha\beta} c_{i\alpha}^\dag c_{j\beta} + \sum_i (\Delta_i c_{i\uparrow}^\dag c_{i\downarrow}^\dag + \text{h.c.}).
\label{eq:Hubb}
\end{align}
Here, $t$ is the hopping element, $\mu$ is the chemical potential, $\vecd_{ij}$ is the vector connecting sites $i$ and $j$, and $\vecm_i$ is the magnetic order parameter on lattice point $i$. The magnetic order parameter on site $i$ is obtained from the staggered antiferromagnetic order parameter $\vecm$ via $\vecm_i =(-1)^{i_a+i_b}\vecm$ with $i_a,i_b$ are labels along the nearest neighbor vector directions $\veca$ and $\vecb$. Moreover, $V_R$ determines the magnitude of the Rashba spin-orbit interaction which is assumed induced from the substrate in both the antiferromagnetic and superconducting region, so that $\hat{\vecn}=-\hat{\vecy}$. Finally, $\Delta_i$ is the superconducting order parameter on site $i$, which is taken as $\Delta_0\e{\pm\ii\phi/2}$ in the left and right superconducting region. The phase difference $\phi$ drives a supercurrent through the system. Ideally the superconducting order parameter should have been solved self-consistently, however this would have drastically increased the computational complexity. To keep the system solvable would have required greatly reducing the system size, so we have here instead assumed a constant $\Delta_i$ inside the superconductors. \\

We consider an (110) interface between the antiferromagnet and a superconductor [see Fig. \ref{fig:model}(a)], so that all the sites closest to the interface have the same magnetization. For an odd number of chains in the antiferromagnet, there will be a net magnetization proportional to
the number of sites on each chain. Such an uncompensated interface with a net magnetization
allows for interesting properties not found in compensated interfaces \cite{andersen_prb_05, enoksen_prb_13, rabinovich_prr_19}. Assuming translational invariance in the $z$-direction, the Fourier-transformed Hamiltonian may be written in compact matrix form $H = H_0 + \frac{1}{2} \sum_k \psi_k^\dag H_k \psi_k$ where $H_0 = -N\sum_{l_x} \mu$ and the basis used is $c_{k\sigma} = [c_{1k\sigma}, \ldots c_{L_xk\sigma}]^\text{T}$, $\psi_k = [c_{k\uparrow}, c_{k\downarrow}, c_{-k\uparrow}^\dag, c_{-k\downarrow}^\dag]$, and $H_k$ is a $4L_x \times 4L_x$ matrix
\begin{align} 
H_k &= \begin{pmatrix}
H_{11k} & H_{12k} & \ldots & H_{1L_xk} \\
\ldots & \ldots & \ldots & \ldots \\
H_{L_x1k}& H_{L_x2k} & \ldots & H_{L_xL_xk} \\
\end{pmatrix}, \notag\\
H_{l_xl_yk} &= \begin{pmatrix}
d_{k\uparrow}& \eta_k & 0 & \Delta \\
\eta_k^* & d_{k\downarrow} & -\Delta & 0 \\
0 & -\Delta^* & -d^*_{-k\uparrow} & -\eta^*_{-k} \\
\Delta^* & 0 & -\eta_{-k} & -d^*{-k\downarrow} \\
\end{pmatrix},\notag\\
d_{k\sigma} &= -2t\cos(\frac{ka}{\sqrt{2}})\delta_+ + m^z_{l_x}(\sigma_z)_{\sigma\sigma} \delta_{l_xl_x'} \notag\\
&+\sqrt{2}\ii a V_R \cos(\frac{ka}{\sqrt{2}})(\sigma_z)_{\sigma\sigma} \delta_- -\mu\delta_{l_xl_x'},\notag\\
\eta_k &= (m_{l_x}^x - \ii m_{l_x}^y)\delta_{l_xl_x'} + \sqrt{2}aV_R\sin(\frac{ka}{\sqrt{2}}) \delta_+,
\end{align}
and $\Delta = \Delta_{l_x}\delta_{l_xl_x'}$ with $\delta_\pm \equiv \delta_{(l_x+1)l_x'} \pm \delta_{(l_x-1)l_x'}$. The indices $l_x$ denote lattice sites along the $x$-direction s.t. even and odd-numbered sites belong to different magnetic sublattices. The matrix $H_k$ is Hermitian and is diagonalized with eigenvalues $\epsilon_{kn}$ and eigenvectors $[u_{k\uparrow n}, u_{k\downarrow n}, v_{k\uparrow n}, v_{k\downarrow n}]^\text{T}$. The expectation value for the current flowing across the junction (in the x-direction) is computed using the Heisenberg equation of motion $\partial_t Q = \ii/\hbar [H,Q]$ and reads:
\begin{align}\label{eq:current}
\langle J_x\rangle &= \frac{4e}{\hbar N} \text{Re}\sum_{kn} \Big( f(\epsilon_{kn}) u^{(l_x+1)\dag}_{kn} (i\hat{t}_k - \hat{V}_R)u^{l_x}_{kn} \notag\\
&-[1-f(\epsilon_{kn})] v^{(l_x+1)\dag}_{kn}(\ii \hat{t}_k + \hat{V}_Rv^{l_x}_{kn})\Big)
\end{align}
with $\hat{t}_k = t\cos(\frac{ka}{\sqrt{2}})$, $\hat{V}_R = \frac{V_R}{\sqrt{2}}[\ii\sin(\frac{ka}{\sqrt{2}})\sigma_x + \cos(\frac{ka}{\sqrt{2}})\sigma_z]$ and $u_{kn}^{l_x}=\big[(u_{k\uparrow n})_{l_x},(u_{k\downarrow n})_{l_x}\big]^T$ with $v_{kn}^{l_x}$ similarly defined, and all sums are understood to only go over positive eigenvalues. The current is to be understood as
the current passing between lattice site $l_x$ and $l_x +1$, and has been normalized to the number of
lattice sites $N$ in the $z$-direction for convenience. The current may be evaluated at an arbitrary point in the antiferromagnetic region as it is conserved.

To compute the ensuing magnetization dynamics of the staggered order parameter induced by a supercurrent, we use the Landau-Lifshitz-Gilbert equation \cite{llg}, which in dimensionless form reads 
\begin{align}
\partial_\tau\vecM_\nu=&-\frac{1}{Nn_\nu}\vecM_\nu\times\vecH_\nu+\alpha\vecM_\nu\times\partial_\tau\vecM_\nu
\label{eq:LLG}
\end{align}
for normalized sublattice magnetization $\vecM_\nu$, where $n_\nu$ is the number of lattice sites of sublattice type $\nu=\{A,B\}$ in the $\vecx$-direction, $\alpha>0$ is the unit-free Gilbert damping and $\tau$ is the dimensionless time coordinate \cite{tau}.
The effective field $\vecH_\nu$ entering the LLG equation is obtained from the derivative of the free energy of the system with respect to the sublattice magnetizations and in dimensionless form reads
\begin{align}\label{eq:effectivefield}
\frac{1}{Nn_\nu} \vecH_\nu &= \sum_{\eta=x,y,z} K_\eta M^\eta_\nu \hat{\boldsymbol{\eta}} - \frac{\Gamma}{Nn_\eta} \frac{1}{t}\frac{\partial F_{JJ}}{\partial \vecM_\nu} \notag\\
&- \frac{2J}{n_\nu} \Big( \sum_{\nu'} n_{\nu'} - 1\Big) \vecM_{\bar{\nu'}},
\end{align}
where $F_{JJ}$ is the free energy of the Josephson junction and we use the hopping element $t$ as the energy scale of the Hubbard model. Here $\hat{\boldsymbol{\eta}}$ is a unit-vector in the $\eta=\{x,y,z\}$-direction, while $K_\eta$, $\Gamma$ and $J$ is unit-free parameters denoting the strength of the cubic magnetic anisotropy, hopping parameter and antiferromagnetic exchange coupling respectively. Note that the Helmholtz free energy is used since we will consider a phase-biased Josephson junction rather than a current-biased one.

$F_{JJ}$ is obtained from the eigenvalues $\epsilon_{n,k}$ of the matrix Hamiltonian $H_k$ through
\begin{align}
F_\text{JJ} = -N\sum_{l_x} \mu - \frac{1}{2}\sum_{n,k} \epsilon_k -\frac{1}{\beta} \sum_{n,k} \text{ln}\Big(1 + \e{-\beta\epsilon_{n,k}}\Big),
\end{align}
where $\beta = 1/k_BT$ is inverse temperature. As seen, the effective field in Eq. (\ref{eq:effectivefield}) couples the magnetization dynamics equations for the two sublattice magnetizations due to the antiferromagnetic exchange $J$. We assume that the magnetic order parameter $\vecm$ in Eq. (\ref{eq:Hubb}) is proportional to the sublattice magnetization, i.e. $\vecm_{l_x}=m\vecM_{l_x}$.

\section{Results}

\subsection{Giant anisotropy in critical supercurrent} 

\begin{widetext}

\begin{figure}[t!]
    \centering
    \includegraphics[width=\columnwidth]{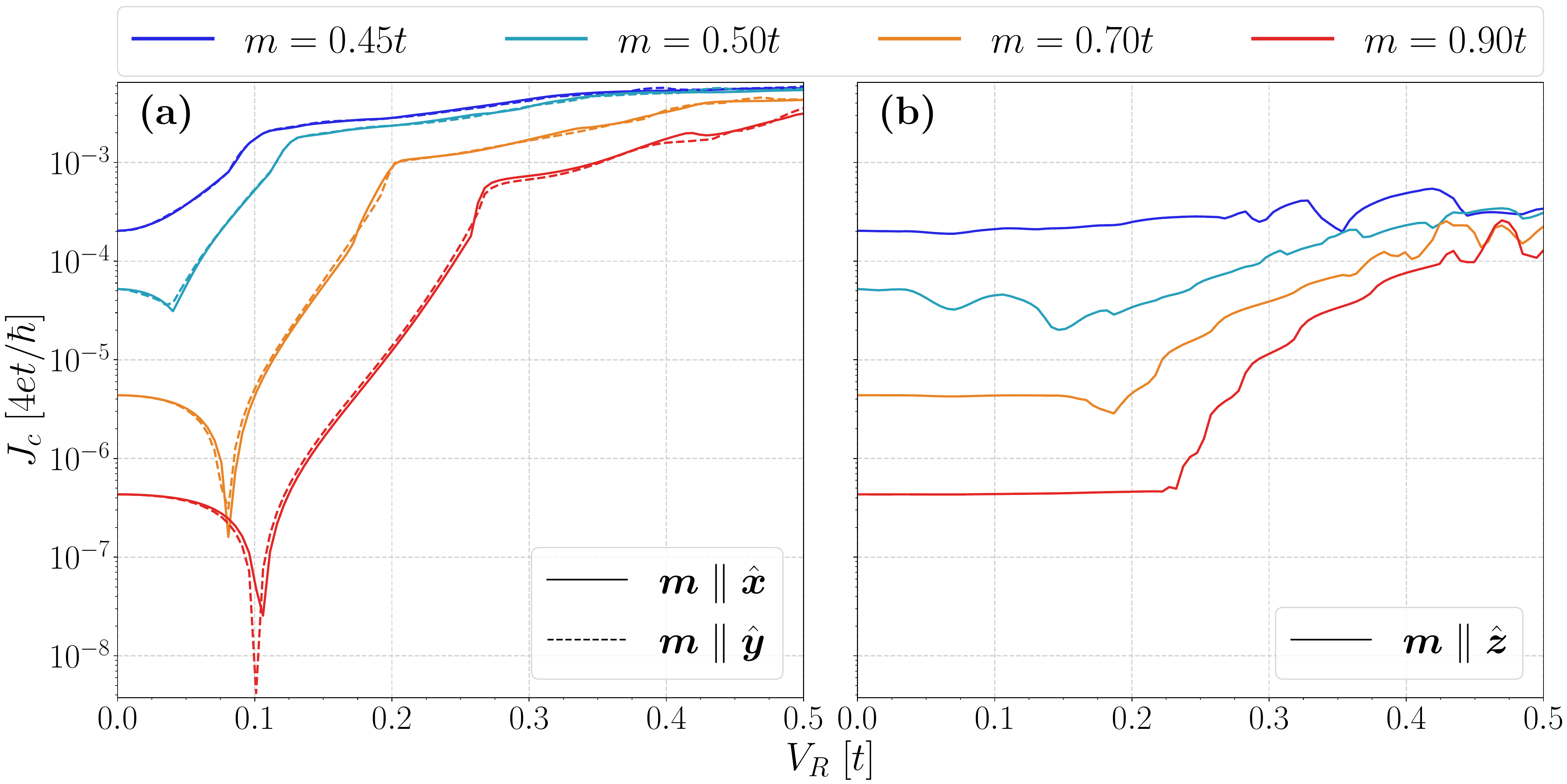}
    \caption{(Color online) The critical current $J_c = |\text{max}_\phi \langle J_x \rangle|$ of the system as a function of the spin-orbit coupling strength $V_R$ using $N=2001$ in the Fourier transform. (a) Staggered order parameter oriented along the $\vecx$ or $\vecy$ direction gives similar results. (b) Staggered order parameter along $\vecz$.
		}
    \label{fig:Jc}
\end{figure}

\end{widetext}

We have computed how the critical supercurrent $J_c = |\text{max}_\phi \langle J_x \rangle|$ changes with the (i) direction of the staggered order parameter and (ii) the magnitude of the spin-orbit coupling in the system. The result is shown in Fig. \ref{fig:Jc}. Due to the possibility of having an anomalous Josephson phase \cite{rabinovich_prr_19}, all values of $\phi \in [-\pi,\pi)$ were checked for each parameter set to obtain the critical value of the current. For numerical calculations we assume that $T=0$, $\mu=0.3t$ and $\Delta_0=0.1t$, with $n_{\text{AF}}=21$ vertical magnetic interlayers and $40$ vertical layers in each superconductor. The chosen value of $\Delta$ is large compared realistic values, in order for the the superconducting coherence length to be comparable to the junction length of the system. Previous theoretical predictions with similar scaling has been shown to compare well with experiments \cite{Black_Schaffer_prb_10, English_prb_2016}. As the magnitude of the antiferromagnetic gap in ordered-Fe-vacancy ternary iron selenides has been experimentally demonstrated to be of order $100 \,\text{meV}$ \cite{Yan_prl_2011} the chosen value of $m/t$ should be reasonable for a poor conductor. The chosen values of $V_R$ is also expected to be within reach experimentally, it has \eg been experimentally demonstrated that Ni(111)/graphene bilayers form a gap of around $200\,\text{meV}$ \cite{Dedkov_prl_2008} due to the Rashba spin-orbit interaction.


Consider first (a) where the staggered order parameter lies either along $\vecx$ or $\vecy$. The initial dip upon increasing the spin-orbit interaction $V_R$ is a $\pi-0$ transition. However, as $V_R$ increases further the critical current is drastically enhanced by several orders of magnitude compared to an antiferromagnetic Josephson junction without spin-orbit coupling $V_R=0$. In panel (b), the staggered order parameter $\vecm$ is oriented along the $\vecz$-direction. In this case, there is no $\pi-0$ transition and the increase in critical current is both delayed to larger values of $V_R$ and is also less dramatic than in panel (a), although still substantial for \eg $m=0.90t$.

\begin{figure}[t!]
    \centering
    \includegraphics[width=\columnwidth]{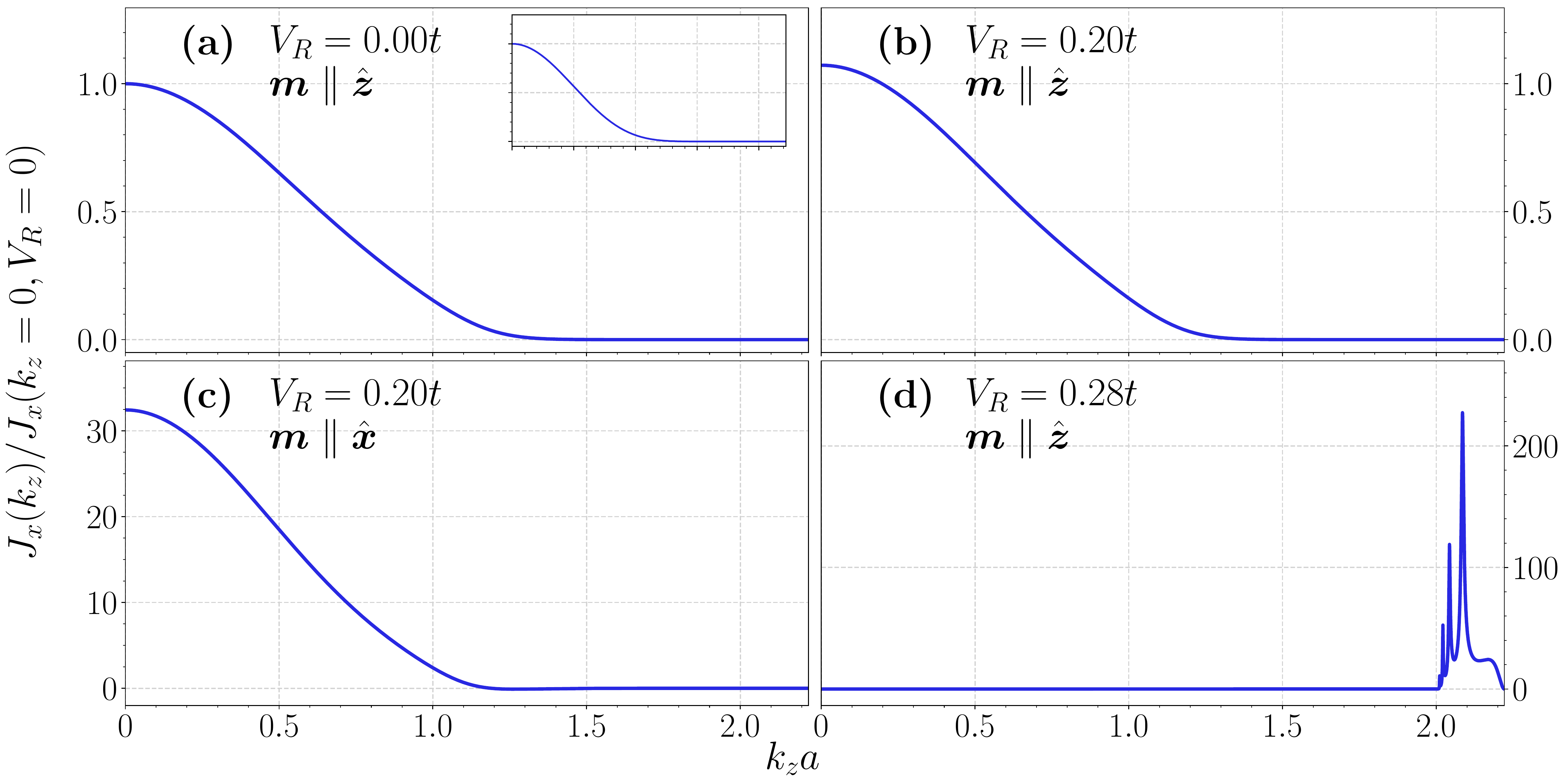}
    \caption{(Color online) The contribution to the critical current for different values of $k_z$ using $m=0.90t$. Each $k_z$ shows the total contribution from $\pm k_z$. The currents have been normalized to the contribution from $k_z=0, V_R=0$. In each panel, the phase difference $\phi$ giving the critical current has been used.
		}
    \label{fig:kz}
\end{figure}

To explain the physical reason this phenomena, we consider the current carried by different transverse modes $k_z$ in the junction, as given by Eq. (\ref{eq:current}). This is shown in Fig. \ref{fig:kz} for several strengths of $V_R$ and orientations of $\vecm$. In the absence of spin-orbit coupling, the current is as expected mostly carried by modes close to normal incidence (small $k_z$) as shown in panel (a). The orientation of the staggered order parameter plays no role for $V_R=0$. Increasing $V_R$, the current is initially still carried by small $k_z$, but is now much larger in the case $\vecm \parallel \vecx$ [(c)] than $\vecm \parallel \vecz$ [(b)], in agreement with Fig. \ref{fig:Jc}. However, when $V_R$ becomes sufficiently large in (d), the current is instead carried mostly by modes away from normal incidence when $\vecm \parallel \vecz$.

\begin{figure}[t!]
    \centering
    \includegraphics[width=\columnwidth]{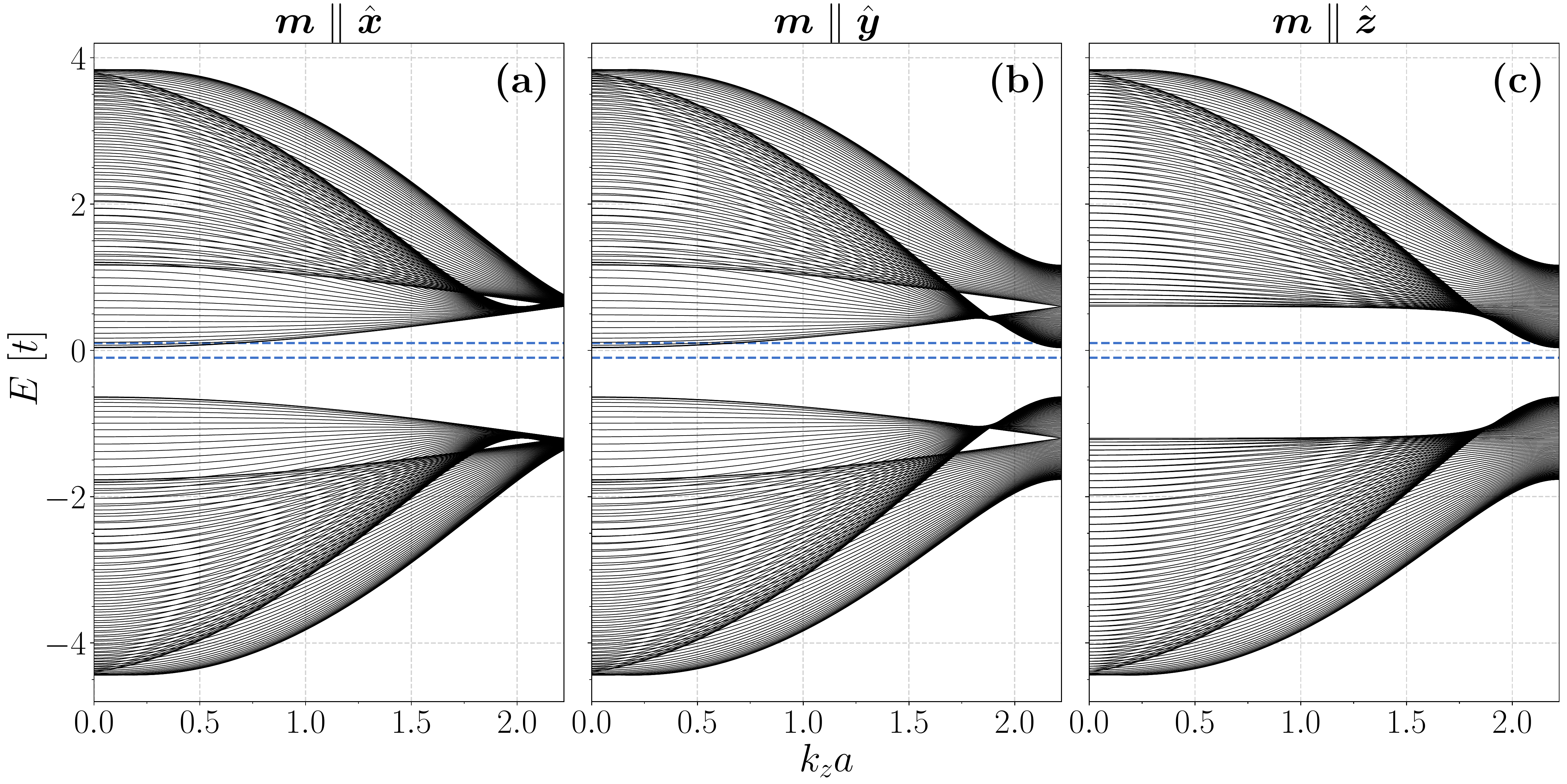}
    \caption{(Color online) The band structure of a spin-orbit coupled antiferromagnet for different orientations of the sublattice magnetization $\vecm$ with $m=0.9t$ and $V_R=0.2t$. Translational invariance is assumed in all directions. Each line is an energy band as a function of $k_z$ with fixed $k_x$. The blue dotted lines indicate the superconducting gap $\pm\Delta$ for ease of comparison. The band structure is symmetric about $k_z=0$. The band structure as a function of $k_x$ for fixed $k_z$ is found by interchanging the plots for $\vecm \parallel \vecx$ and $\vecm \parallel \vecz$. 
		}
    \label{fig:band}
\end{figure}

To understand this, we finally need to look at the band structure of the spin-orbit coupled antiferromagnet. Assuming translational invariance in all directions, we have computed the band structure in Fig. \ref{fig:band} for $m=|\vecm|=0.9t$ and $V_R=0.2t$. 
The first important observation is that the spin-orbit coupling causes the bandgap in the antiferromagnet to narrow for specific values of $\vk$. In fact, for sufficiently large $V_R$, the bandgap can close entirely. The key effect which lies behind the giant anisotropy of the critical supercurrent is that the $\vk$-values at which the bandgap narrows depends on the orientation of $\vecm$, as seen in Fig \ref{fig:band}. For $\vecm \parallel \vecx$, the gap narrows for small $k_z$. In contrast, for $\vecm \parallel \vecz$, the gap narrows for large $k_z$. For $\vecm \parallel \vecy$, the gap narrows for both small and large $k_z$. 


This effect can be understood from the combination of spin-orbit interaction and the antiferromagnetic order. Spin-orbit interaction couples sites on opposite sublattices, where spins prefer anti-parallel alignment along $\boldsymbol{m}$ due to the antiferromagnetic order. Normally in a translationally invariant system, spin-orbit interaction prefers the spin to be aligned orthogonally to both $\hat{\boldsymbol{n}}$ and $\boldsymbol{k}$. However for alternating spins, the spin-orbit interaction instead couples to spin-orientations in the plane spanned by $\hat{\boldsymbol{n}}$ and $\boldsymbol{k}$. Thus for small $k_z$ the spin-orbit interaction mainly couples to the $x$- and $y$-components of the staggered order parameter, which can cause a narrowing of the gap. For $\boldsymbol{m}\parallel\hat{\boldsymbol{z}}$ the spin-orbit coupling at $k_z=0$ does not couple to the antiferromagnetic order, and as such cannot narrow the gap.

The critical supercurrent supported by the junction will clearly  depend on the magnitude of the band gap in the antiferromagnet, since it becomes exponentially damped with increasing gap. Since we showed above (Fig. \ref{fig:kz}) that the supercurrent is primarily carried by modes with small $k_z$, the narrowing gap with increasing $V_R$ for $\vecm \parallel \vecx,\vecy$ then explains the large increase in critical current seen in Fig. \ref{fig:Jc}(a). In contrast, the narrowing of the gap for large $k_z$ in Fig. \ref{fig:band}(c) changes slowly with increasing $V_R$. This causes the supercurrent to only depend weakly on $V_R$ for $\vecm \parallel \vecz$ until a critical value of $V_R$ is reached where the gap is closed for large $k_z$, in agreement with Fig. \ref{fig:Jc}(b). This closing of the gap in the case $\vecm \parallel \vecz$ for large values of $k_z$ explains both the $k$-dependence in Fig. \ref{fig:kz}(d) and why the critical current sharply starts to increase at a critical value of $V_R$ i Fig. \ref{fig:Jc}(b).

\begin{figure}[b!]
    \centering
    \includegraphics[width=\columnwidth]{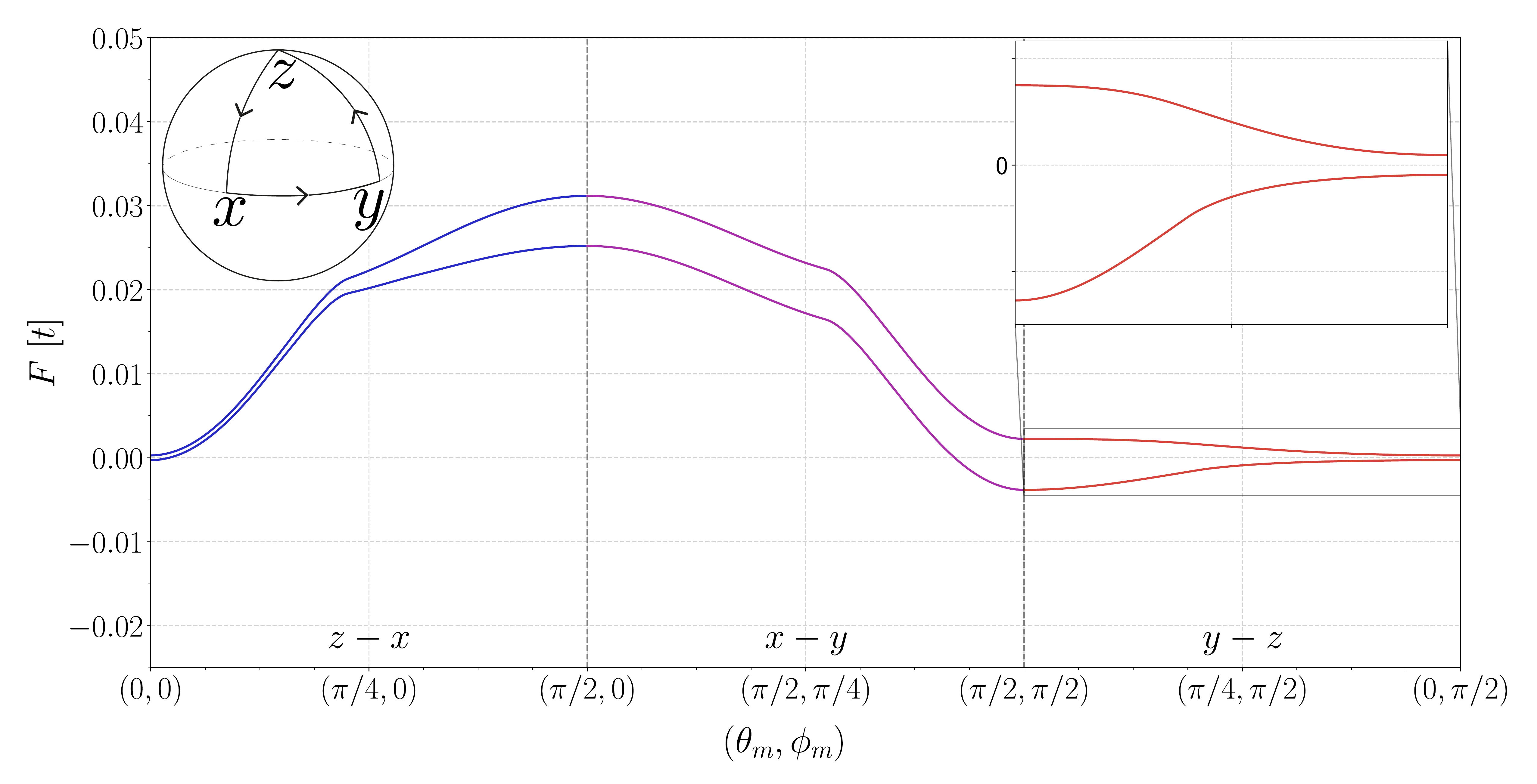}
    \caption{(Color online) The free energy of a Josephson junction with $m=0.45t$, $V_R=0.10t$ and $N=601$ as the staggered order parameter $\vecm$ is rotated around the triangle $z \to x \to y \to z$ on the sphere in the inset. Free energy including a magnetic anisotropy $F_\text{anis} = -n_\text{AF} K(m_y^2 - m_x^2)/2$ with $n_\text{AF}Km^2/2 = 0.043t$. The upper line represents the free energy for $\phi=\pi$ whereas the lower line is for $\phi=0$.}
    \label{fig:free_energy}
\end{figure}

The main result in the above analysis is thus that for a fixed value of the spin-orbit coupling $V_R$, the critical current can be tuned several orders in magnitude by rotating the staggered order parameter $\vecm$. We note that a similar effect could also exist in junctions with insulating ferromagnets, the only difference between this model and that of a ferromagnetic junction is the alternating direction of the order parameter $\vecm$. However, in a ferromagnetic junctions much larger values of $m$ is required to open a gap in the ferromagnet, which which could change the behaviour of the system. The momentum dependence is also expected to differ as neighboring spins prefer parallel instead of anti-parallel orientations, changing the effect of the spin-orbit interaction.

\subsection{Staggered magnetization dynamics in phase-biased junctions}
The above analysis showed that the supercurrent is highly sensitive to the orientation of the antiferromagnetic order parameter $\vecm$. This suggests the existence of the reciprocal effect: a rotation of the sublattice magnetization $\vecM$ caused by a phase-difference across the junction. To determine if this is possible, we start by analyzing the free energy of the system and how it depends on the orientation of the sublattice magnetizations. The total free energy of the junction including an staggered order parameter anisotropy term $F_\text{anis} = -n_\text{AF} K(m_y^2 - m_x^2)/2$ is shown in Fig. \ref{fig:free_energy}. The system minimizes the free energy for a specific orientation of the order parameter $\vecm$ based on the non-trivial competition between the superconducting proximity effect, band-structur effects, and magnetic anisotropies in the system. The key observation is seen in the inset of Fig. \ref{fig:free_energy}: the orientation of the staggered order $\vecm$ which minimizes $F$ depends on the phase difference across the junction. For $\phi=0$, the preferred orientation is $\vecm \parallel \vecy$, whereas $\phi=\pi$ favors $\vecm \parallel \vecz$. The phase of the Josephson junction is controllable by integrating the junction into a superconducting ring threaded by a weak magnetic flux.

\begin{widetext}

\begin{figure}[t!]
    \centering
    \includegraphics[width=\columnwidth]{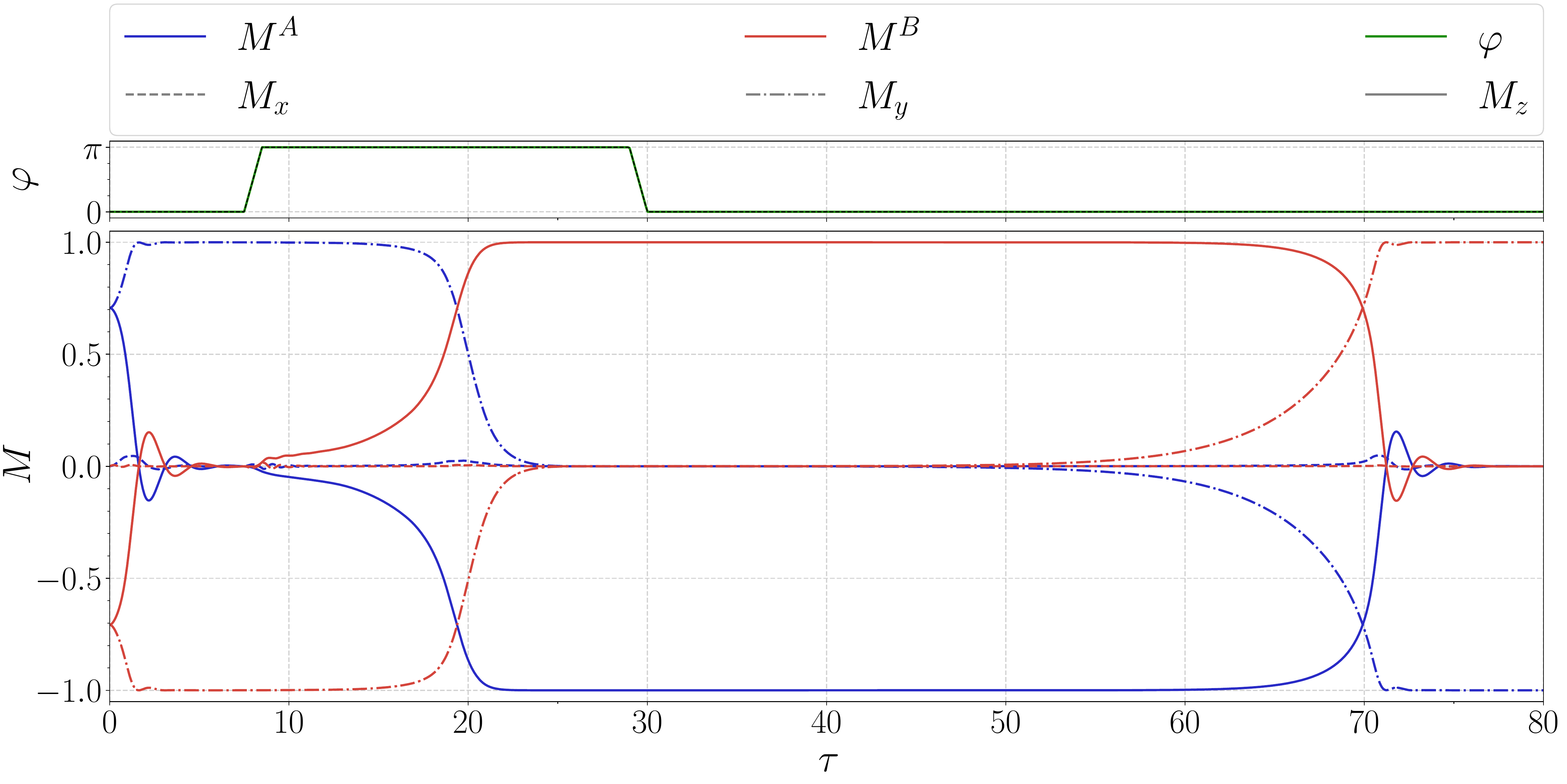}
    \caption{(Color online) The magnetization dynamics of the staggered order parameter when the phase is locked alternatively to $0$ or $\pi$ (upper panel). As seen, magnetization switching where the staggered order flips 90 degrees is attainable. The parameters used are $m=0.45t$, $V_R=0.10t$ and $N=101$ for the Josephson junction while $\tilde{\alpha} = 0.06, K_x=-1$, $K_y=1$, $K_z=0$, $\Gamma=n_\text{AF}/0.086$ and $J=1$.}
    \label{fig:magdyn}
\end{figure}

\end{widetext}

We have solved the LLG equations Eqs. (\ref{eq:LLG}) and (\ref{eq:effectivefield}) numerically with zero phase difference $\phi=0$ as initial condition. As expected from the free energy analysis above, the system attempts to orient $\vecM$ along $\vecy$. However, when switching $\phi$ to $\pi$, $\vecM \parallel \vecz$ is energetically preferred, causing a reorientation of the sublattice magnetization. The anomalous phase difference in the junction plays no particular role in this dynamic behavior, since the chosen parameters ensure that such an anomalous phase is essentially absent for all orientations of $\vecm$. Switching the phase back to $\phi=0$, $\vecM$ switches back to $\parallel \vecy$. It is interesting to note that the speed at which the transitions $\vecy\to\vecx$ and $\vecz \to -\vecy$ for the staggered order $\vecM$ take place are different. This is due to the free energy landscape in the $\phi-\vecM$ plane, which has a steepness that is not identical for $\phi=0$ and $\phi=\pi$, as seen in Fig. \ref{fig:free_energy}.

\section{Conclusion}
Our results enable both Néel order parameter control via the superconducting phase and a magnetic transistor-like functionality for a supercurrent with a critical current varying by several orders of magnitude when rotating the Néel order parameter. These findings reveal an interesting versatility of antiferromagnetic Josephson junctions resulting from the interplay of staggered and phase-coherent order which can be tested experimentally.

\begin{acknowledgments}
We thank M. D. Hansen and I. Bobkova for useful discussions. We acknowledge funding from the Research Council of Norway via its Centres of Excellence funding scheme, project number 262633.
\end{acknowledgments}

\appendix

\section{Effect of self-consistency in $\Delta$}

\begin{figure}[b!]
    \centering
    \includegraphics[width=\columnwidth]{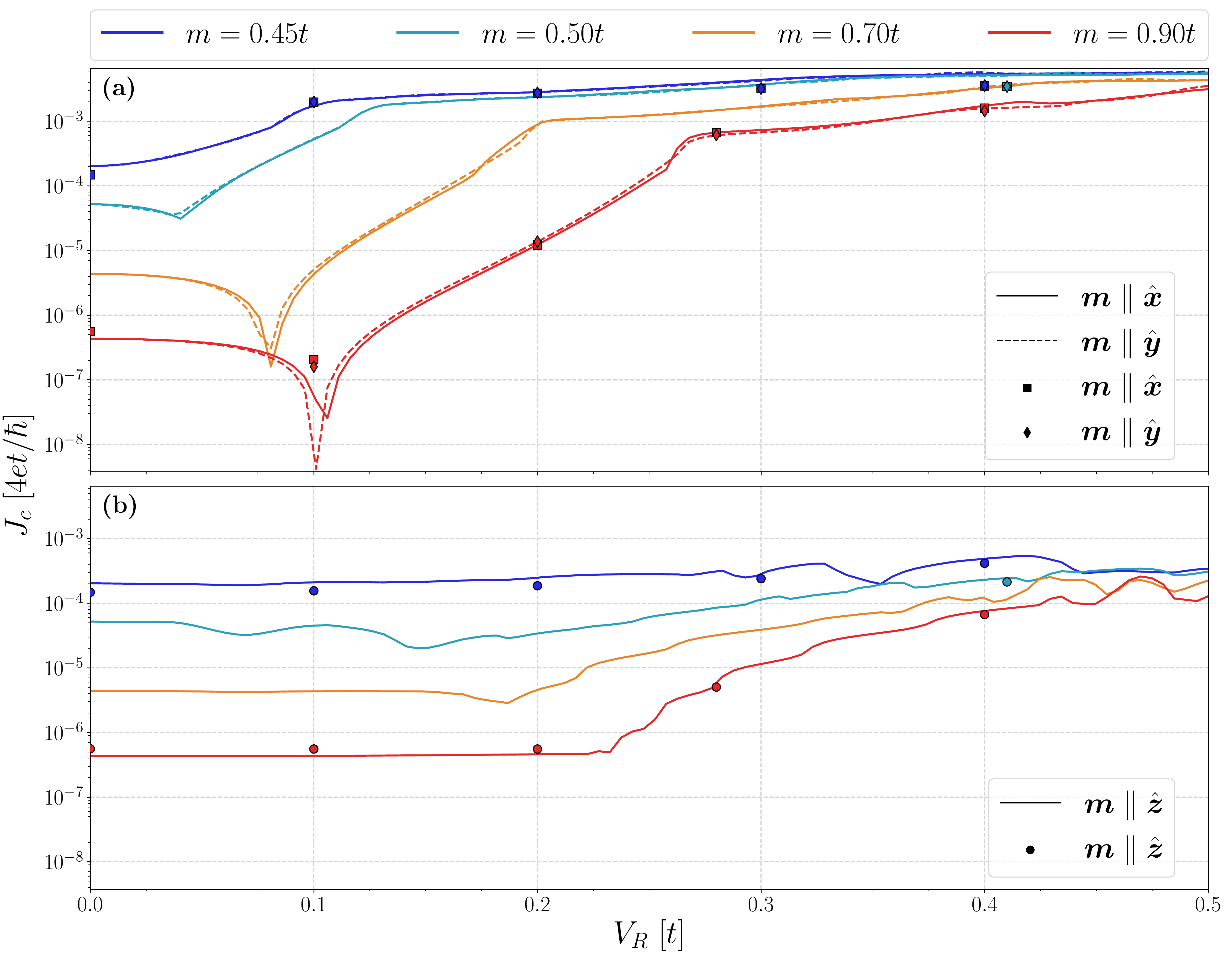}
    \caption{(Color online) Comparison of selfconsistent and non-selfconsistent solution in the superconducting order parameter for the magnetization dynamics of the staggered order parameter (see also Fig. \ref{fig:Jc}). The superconducting coupling constant is set to $V=1.395t$.}
    \label{fig:CPRSC}
\end{figure}

Selfconsistency in the order parameters \cite{andersen_prl_2006} was not considered in the main text. This is a common approximation due to the large increase in computational resources required to study a system selfconsistency. However, to make sure that our results do not change significantly when doing the computations selfconsistency, we show in Fig. \ref{fig:CPRSC} a comparison between selfconsistent and non-selfconsistent calculations for the superconducting order parameter. Due to the time-consumption required for the selfconsistent solution, a smaller number of $V_R$ values have been considered compared to the non-selfconsistent solution. As seen, the two procedures match qualitatively and are comparable quantitatively. Moreover, since the superconducting gap is the smallest energy scale in the problem, it is expected to be the most strongly affected by selfconsistency. Therefore, we have not considered the effect of selfconsistency on the antiferromagnetic order parameter, since we do not expect any substantial change in the results.

\end{document}